PAPER • OPEN ACCESS

# Solving Klein's paradox

View the article online for updates and enhancements.

# Journal of Physics Communications

PAPER

# Solving Klein's paradox

Huai-Yu Wang

Department of Physics, Tsinghua University, Beijing 100084, People's Republic of China

E-mail: wanghuaiyu@mail.tsinghua.edu.cn

**Abstract**
We figure out the famous Klein's paradox arising from the reflection problem when a Dirac particle encounters a step potential with infinite width. The key is to piecewise solve Dirac equation in such a way that in the region where the particle's energy $E$ is greater (less) than the potential $V$, the solution of the positive (negative) energy branch is adopted. In the case of Klein–Gordon equation with a piecewise constant potential, the equation is decoupled to positive and negative energy equations, and reflection problem is solved in the same way. Both infinitely and finitely wide potentials are considered. The reflection coefficient never exceeds 1. The results are applied to discuss the transmissions of particles with no mass or with very small mass.

## 1. Introduction

Shortly after Dirac equation was proposed, Klein came up with his famous paradox [1]. The main content of this paradox was that when a relativistic particle encountered an infinitely wide potential barrier with height exceeding a certain value, the reflection coefficient would be larger than 1 [2]. There have been investigations on this paradox from different viewpoints [3–7]. For Klein–Gordon equation, there was the same paradox [3, 8–10]. One interpretation of this result was to use Dirac's concept of electron sea full of negative energy electrons, which thought that electrons, as well as positrons, were extracted from the sea in a sufficiently strong field [4, 10–13]. For Klein–Gordon equation which described the motion of particles with spin zero, it was also believed that antiparticles were produced under strong field [3]. However, this interpretation was a plausible one since it was unable to calculate the amount of the electrons produced and the energy conservation was hardly to clarify. Anyway, the problem has remained open. The present work intends to ultimately solve the paradox.

It is noticed that the paradox merged when a relativistic particle's energy was less than the height of the potential barrier it was in. That prompts us to inspect the relationship between a particle's energy and potential.

In this work, we consider the stationary motion of a particle with energy $E$ in one-dimensional space.

Dirac equation is that

$$\begin{pmatrix} mc^2 + V & -ic\hbar d/dx \\ -ic\hbar d/dx & -mc^2 + V \end{pmatrix}\psi = E\psi. \tag{1.1}$$

First, let us see the case when potential is absent. It is solved from equation (1.1) that a free particle is of energy

$$E_{(+)} = \sqrt{m^2c^4 + c^2p^2} \tag{1.2a}$$

and

$$E_{(-)} = -\sqrt{m^2c^4 + c^2p^2}. \tag{1.2b}$$

There are two branches. Each branch has its own eigen function: that belonging to $E_{(+)}$ ($E_{(-)}$) is denoted as $\psi_{(+)}$ ($\psi_{(-)}$). Please note that for free particles, both $\psi_{(+)}$ and $\psi_{(-)}$ are plane waves.

Then we consider that there is a potential $V$. The energies should be respectively

$$E_{(+)} = \sqrt{m^2c^4 + c^2p^2} + V > V \tag{1.3a}$$

and

$$E_{(-)} = -\sqrt{m^2c^4 + c^2p^2} + V < V. \tag{1.3b}$$

Still, each branch has its own eigen function. However, the properties of the solutions when there is a discontinuous potential should be stressed. It is seen that $E_{(+)} > V$, which means that the eigen function $\psi_{(+)}$ belonging to $E_{(+)}$ only applies to regions where $E > V$. On the other hand, $E_{(-)} < V$, which means that the eigen function $\psi_{(-)}$ belonging to $E_{(-)}$ only applies to regions where $E < V$.

Dirac equation describes the motion of particles with spin 1/2. Particles with spin 0 obey Klein–Gordon equation.

$$(E - V)^2 \psi = \left(m^2c^4 - c^2\hbar^2 \frac{d^2}{dx^2}\right)\psi. \tag{1.4}$$

When potential is absent or the potential is piecewise constant, the equation can be rewritten as

$$(E - H_{(-)})(E - H_{(+)})\psi = 0, \tag{1.5}$$

where we have denoted that

$$H_{(\pm)} = \pm H_0 + V, \quad H_0 = \sqrt{m^2c^4 - c^2\hbar^2 \frac{d^2}{dx^2}}. \tag{1.6}$$

Now, let us impose a restriction that

$$(E - H_{(+)})\psi = 0. \tag{1.7a}$$

The two factors in equation (1.5) can also be exchanged and we impose another restriction that

$$(E - H_{(-)})\psi = 0. \tag{1.7b}$$

Thus, in the case of piecewise constant potential, Klein–Gordon equation is decoupled into two. Hereafter equations (1.7a) and (1.7b) are called decoupled Klein–Gordon equations. For a free particle, the eigenvalue of equation (1.7a) is (1.2a), and that of (1.7b) is (1.2b). Each has its own plane wave as eigen function.

When there is a potential, then the energy of equation (1.7a) is (1.3a), and that of (1.7b) is (1.3b). Please note that in the stationary motion, a particle's energy $E$ remains unchanged everywhere, while potential may depends on coordinate. Since equation (1.3a) shows $E > V$, equation (1.7a) applies in regions where $E > V$. This equation is called decoupled Klein–Gordon equation of positive energy branch. Similarly, equation (1.7b) is applicable in regions where $E < V$, and it is called decoupled Klein–Gordon equation of negative energy branch.

The clarification of the relationship between a particle's energy and potential it is in helps one to correctly use the equation and corresponding wave function for both Dirac equation and decoupled Klein–Gordon equations.

With the preparation above, we are ready to solve Klein's paradox. In any case, the reflection coefficient cannot be larger than 1. No extraordinary concepts are needed, such as a far-fetched interpretation of the excitation of a pair of positron and electron from the electron sea.

The potential barrier in Klein's paradox was of infinite width. For a potential barrier with finite width, a particle is able to penetrate the barrier even in the case $E < V$, the famous tunnel effect.

In this work, we will also evaluate the reflection coefficient of one-dimensional square potential barrier. For both spin 1/2 and spin0 particles, the reflection coefficient cannot be greater than 1. It will be shown that particles penetrate potential barrier much easier than expected.

This paper calculates the reflection coefficient of one-dimensional step potential. In section 2, we deal with infinitely wide potential barrier. The main aim is to figure out Klein's paradox. The reflection coefficient of Klein–Gordon equation is also evaluated. In section 3, we treat the case of a square potential with finite width. Both Dirac equation and Klein–Gordon equation are considered. We also discuss the cases where a particle's mass is zero or very small. Section 4 is our conclusion.

Relativistic quantum mechanics equations, Dirac equation and Klein–Gordon equation, have positive and negative energy branches. The present work shows the necessity to use the solutions belonging to the negative energy branch in regions where a particle's energy $E$ is less than potential $V$. We have found that when a particle did low momentum motion, the negative energy branch should still remain [14]. This, as a matter of fact, can be easily obtained by expanding the square root of equation (1.3b) and making low momentum approximation. Thus, for a low momentum particle, when its energy $E > V$ it followed Schrödinger equation, while when $E < V$, it followed negative kinetic energy Schrödinger equation developed in [14]. Therefore, the positive and negative energy branches existed for both relativistic and low momentum motions. The author's primary motivation was to treat the two energy branches on an equal footing and to explore the physical meaning of the solutions of the negative energy branch and the applications of the solutions. Reference [14] was the author's first paper on this topic. In appendix D of [14], some works in undertaking and to be done were listed. The present paper concerns point 2 there.

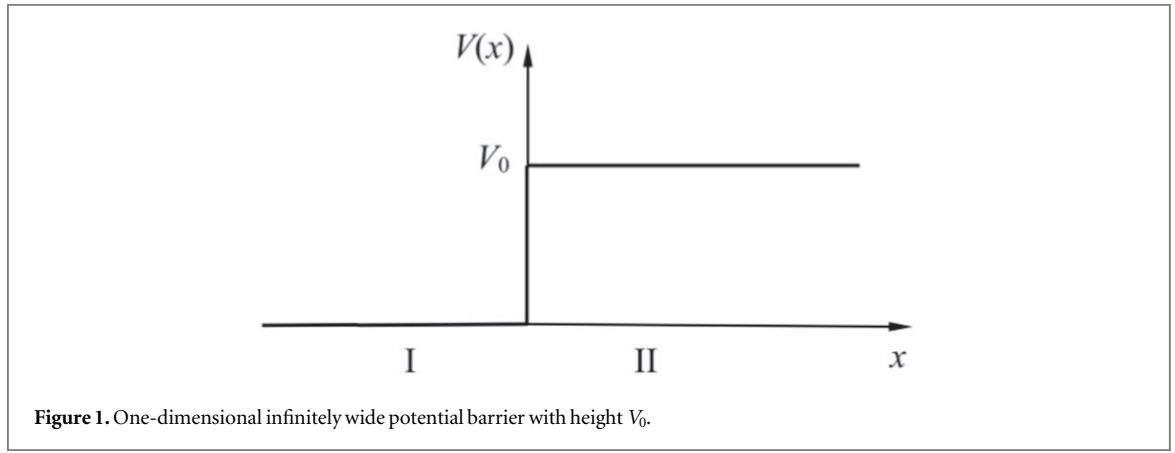

**Figure 1.** One-dimensional infinitely wide potential barrier with height $V_0$.

## 2. Infinitely wide potential barrier

In this section, we consider the issue of infinitely wide potential barrier. The potential is as follows.

$$V(x) = \begin{cases} 0, & x \leqslant 0 \\ V_0, & x > 0 \end{cases} \tag{2.1}$$

Please see figure 1. A particle with energy $E$ and momentum $q$ is incident from left and moves rightwards. Let us investigate the solutions of Dirac equation.

### 2.1. Klein's paradox
In the region $x < 0$, the energy-momentum relation of the incident particle is

$$E^2 = q^2 c^2 + m^2 c^4. \tag{2.2}$$

In the barrier region $x > 0$, assuming that its momentum is $p$, we have

$$(E - V_0)^2 = p^2 c^2 + m^2 c^4. \tag{2.3}$$

The wave function in region $x < 0$ is written as

$$\psi_{\mathrm{I}} = \begin{pmatrix} qc \\ E - mc^2 \end{pmatrix} e^{iqx/\hbar} + B \begin{pmatrix} -qc \\ E - mc^2 \end{pmatrix} e^{-iqx/\hbar} \tag{2.4}$$

with $B$ being reflection amplitude. Similarly, the wave function in region $x > 0$ was written as

$$\psi_{\mathrm{II}} = F \begin{pmatrix} pc \\ E - V_0 - mc^2 \end{pmatrix} e^{iqx/\hbar} \tag{2.5}$$

with $F$ being transmission amplitude.

At boundary $x = 0$, the wave function should be continuous, so that

$$qc(1 - B) = Fpc, \tag{2.6a}$$

and

$$(E - mc^2)(1 + B) = F(E - V_0 - mc^2). \tag{2.6b}$$

Let

$$\alpha = \frac{1 + B}{1 - B} = \frac{\sqrt{E^2 - m^2 c^4}}{\sqrt{(E - V_0)^2 - m^2 c^4}} \frac{E - V_0 - mc^2}{E - mc^2}. \tag{2.7}$$

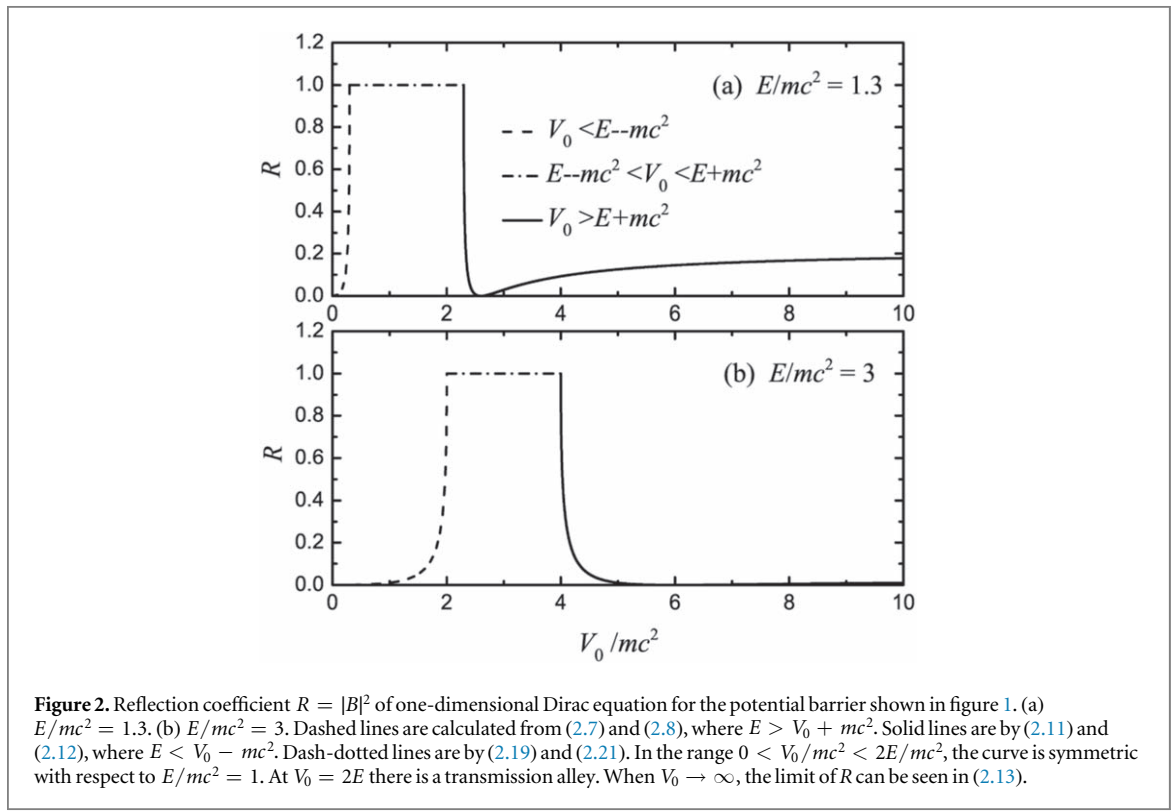

**Figure 2.** Reflection coefficient $R = |B|^2$ of one-dimensional Dirac equation for the potential barrier shown in figure 1. (a) $E/mc^2 = 1.3$. (b) $E/mc^2 = 3$. Dashed lines are calculated from (2.7) and (2.8), where $E > V_0 + mc^2$. Solid lines are by (2.11) and (2.12), where $E < V_0 - mc^2$. Dash-dotted lines are by (2.19) and (2.21). In the range $0 < V_0/mc^2 < 2E/mc^2$, the curve is symmetric with respect to $E/mc^2 = 1$. At $V_0 = 2E$ there is a transmission alley. When $V_0 \to \infty$, the limit of $R$ can be seen in (2.13).

The reflection amplitude was expressed by

$$B = \frac{\alpha - 1}{\alpha + 1}. \tag{2.8}$$

It is then easy to calculate the reflection coefficient $R = |B|^2$ and transmission coefficient $T = |F|^2 = 1 - R$. As $V_0 = 0$, the particle certainly transmits totally. When $V_0$ increases from zero to $E - mc^2$, $R$ is from zero to 1, see dashed lines in figure 2. However, as $V_0$ increases further, the result $R > 1$ will appear, which means that particles seem generated and transmission coefficient is negative. Especially, it can be reckoned from (2.7) and (2.8) that as $V_0 \to \infty$, $\alpha \to -1$ so that $R \to \infty$, a ridiculous result. This was the famous Klein's paradox.

This paradox was interpreted by electron generation from Dirac Sea, which was not convincing as we have mentioned in Introduction.

We are going to figure out the paradox in the next subsection.

### 2.2. Solving Klein's paradox

In fact, the wave function (2.5) merely applies to the case of $V_0 < E - mc^2$. The case of $V_0 > E - mc^2$ is complicated. We have to discuss it by dividing the height $V_0$ into three ranges.

Let us first discuss the range $V_0 > E + mc^2$. Because the particle's energy is less than the barrier height, in the barrier region $x > 0$, the wave function should be $\psi_{(-)}$ belonging to equation (1.3b). So it is written as

$$\psi_{II} = F \begin{pmatrix} -E + V_0 + mc^2 \\ pc \end{pmatrix} e^{ipx/\hbar}. \tag{2.9}$$

In region $x < 0$, equation (2.4) still applies. So we have (2.4) and (2.9) available. At the boundary $x = 0$, the wave function should be continuous. It follows that

$$qc(1 - B) = F(-E + V_0 + mc^2) \tag{2.10a}$$

and

$$(1 + B)(E - mc^2) = Fpc. \tag{2.10b}$$

Let

$$\beta = \frac{\sqrt{(E - mc^2)(V_0 - E + mc^2)}}{\sqrt{(E + mc^2)(V_0 - E - mc^2)}}. \tag{2.11}$$

The reflection amplitude can be evaluated by

$$B = \frac{\beta - 1}{\beta + 1}. \tag{2.12}$$

The reflection coefficient is $R = |B|^2$ and transmission coefficient is $T = |F|^2 = 1 - R$. When $V_0 < 2E$, exchange of $V_0 - E \Leftrightarrow E - V_0$ makes equation (2.7) become (2.11) and vice versa. Numerical results are plotted by solid lines in figure 2. It is shown that as $V_0$ increases from $E + mc^2$, reflection coefficient decreases. As $V_0 = 2E, R = 0$, which means that the particle has a total transmission. We say that there is a transmission alley around this barrier height. The transmission alley has a clear physical meaning: as $V_0 = 2E, p = q$, see equations (2.2) and (2.3). That is to say, the value of the momentum of the particle does not alter before and after it enters the barrier. This situation is equivalent to that it is not scattered, and is similar to the case of 'complete impedance matching' in electromagnetic materials.

As $V_0$ rises further, $R$ also rises from zero. When $V_0 \to \infty$,

$$R \to \left(\frac{mc^2}{E + \sqrt{E^2 - m^2c^4}}\right)^2 < 1. \tag{2.13}$$

In this limit, the reflection coefficient approaches a fixed value that is less than 1. This fixed value decreases with the energy $E$ increasing, as well as the ratio $mc^2/E$ decreasing. That is to say, under a strong potential barrier, the transmission coefficient rises as the particle's energy does.

Now we turn to the barrier height within range $E - mc^2 < V_0 < E + mc^2$, which actually means that

$$V_0 - mc^2 < E < V_0 + mc^2. \tag{2.14}$$

This clearly demonstrates that the particle's energy is within the energy gap of relativistic quantum mechanics. In this case, because $|V_0 - E| < +mc^2$,

$$p^2c^2 = -q^2c^2 = (V_0 - E)^2 - m^2c^4 < 0, \tag{2.15}$$

which shows that momentum is an imaginary number. We are therefore aware of that within the relativistic energy gap, the momentum is imaginary.

We have to put down the wave functions of one-dimensional Dirac equation when energy is within the range (2.14). They are as follows.

When $E - mc^2 < V_0 < E$, the case belonging to $E > V$,

$$\psi : \begin{pmatrix} ikc \\ E - V_0 - mc^2 \end{pmatrix} e^{-kx/\hbar}, \begin{pmatrix} -ikc \\ E - V_0 - mc^2 \end{pmatrix} e^{kx/\hbar}. \tag{2.16}$$

When $E < V_0 < E + mc^2$, the case belonging to $E < V$,

$$\psi : \begin{pmatrix} E - V_0 + mc^2 \\ -ikc \end{pmatrix} e^{kx/\hbar}, \begin{pmatrix} E - V_0 + mc^2 \\ ikc \end{pmatrix} e^{-kx/\hbar}. \tag{2.17}$$

Having the wave functions (2.16) and (2.17) available, we are now ready to evaluate the reflection coefficient when the particle's energy is in the gap.

When $E - mc^2 < V_0 < E$, the wave function in region $x > 0$ should be chosen from (2.16). It is

$$\psi_{II} = F \begin{pmatrix} ikc \\ E - V_0 - mc^2 \end{pmatrix} e^{-kx/\hbar}. \tag{2.18}$$

The simultaneous equations are (2.4) and (2.18). The condition of continuity at $x = 0$ leads to

$$\beta = \frac{1 - B}{1 + B} = -i\frac{\sqrt{(E - mc^2)(E - V_0 + mc^2)}}{\sqrt{(E + mc^2)(V_0 - E + mc^2)}}, \tag{2.19}$$

which is a purely imaginary number. Then the reflection amplitude $B$ is calculated, and the reflection coefficient is $R = |B|^2 = 1$.

When $E < V_0 < E + mc^2$, the wave function in region $x > 0$ should be chosen from (2.17) and it is

$$\psi_{II} = F \begin{pmatrix} E - V_0 + mc^2 \\ ikc \end{pmatrix} e^{-kx/\hbar}. \tag{2.20}$$

The simultaneous equations are (2.4) and (2.20). The condition of continuity at $x = 0$ leads to

$$\beta = \frac{1-B}{1+B} = -\mathrm{i}\frac{\sqrt{(E-mc^2)(V_0-E+mc^2)}}{\sqrt{(E+mc^2)(E-V_0+mc^2)}}, \qquad (2.21)$$

which is again an imaginary number. Therefore, the reflection coefficient is again 1.

It is seen that because in the range $E - mc^2 < V_0 < E + mc^2$, the particle's momentum in the barrier is imaginary and the wave function exponentially decays, it is always totally reflected. This is embodied in figure 2 as a platform of $R = 1$. This platform is called 'energy gap reflection platform'.

It is seen from above discussion that in any case, the reflection coefficient cannot be larger than 1.

We therefore conclude that we have completely figured out Klein's paradox without need of imaging exciting electrons from Dirac Sea. The key is to take the solution $\psi_{(-)}$ in the region where potential is higher than particle's energy, equation (1.3b). There was a work mentioning the negative solution in dealing with Klein's paradox, but it seemed that the wave function of the negative energy solution was not employed [11].

Inside the barrier, if the particle's energy is within the relativistic energy gap, the wave function is in exponential form, see equations (2.18) and (2.20), while if it is not, the particle propagates in the form of plane wave, see equation (2.9).

### 2.3. The case of Klein–Gordon equation

Klein raised his paradox when he dealt with Dirac equation describing the relativistic motion of particles with spin 1/2. For Klein–Gordon equation, there was also Klein's paradox, as mentioned in introduction, although this equation, the wave function of which has only one component, is simpler than Dirac equation.

Let use consider the case of $E < V_0$. Intuitively, because the energy is less than the barrier height, the wave in the barrier would be thought as exponentially decaying one. This requires that momentum must be imaginary. However, according to the relativistic energy-momentum relationship of Klein–Gordon equation, equation (2.3), the momentum is not imaginary when $E < V_0 - mc^2$. Moreover, even within the range $E - mc^2 < V_0 < E$ when the barrier height is less than the particle's energy, the momentum has to be imaginary. This is because energy $E$ satisfying $|V_0 - E| < +mc^2$ is within relativistic energy gap, as shown by (2.14).

The correct treatment of Klein–Gordon equation for the step potential problem should also be presented. This goal can be achieved by use of the decoupled Klein–Gordon equations, equation (1.7): depending on the relationship between the particle's energy $E$ and barrier height $V_0$, we choose corresponding wave function $\psi_{(+)}$ or $\psi_{(-)}$.

In region $x < 0$, the wave function is

$$\psi_\mathrm{I} = \mathrm{e}^{\mathrm{i}qx/\hbar} + B\mathrm{e}^{-\mathrm{i}qx/\hbar} \qquad (2.22)$$

and the energy is expressed by

$$E^2 = q^2c^2 + m^2c^4. \qquad (2.23)$$

In region $x > 0$, we have to consider three ranges of $V_0$ value just as in the case of Dirac equation. When $V_0 < E - mc^2$ and $V_0 > E + mc^2$, we have

$$\psi_\mathrm{II} = F\mathrm{e}^{\mathrm{i}px/\hbar} \qquad (2.24)$$

and

$$(E-V_0)^2 = p^2c^2 + m^2c^4. \qquad (2.25)$$

The evaluated reflection coefficient is

$$R = \left[\frac{\sqrt{E^2 - m^2c^4} - \sqrt{(E-V_0)^2 - m^2c^4}}{\sqrt{E^2 - m^2c^4} + \sqrt{(E-V_0)^2 - m^2c^4}}\right]^2. \qquad (2.26)$$

In these two ranges, all the formulas are the same. This equation remains invariant under the exchange of $V_0 - E \Leftrightarrow E - V_0$.

When $E - mc^2 < V_0 < E + mc^2$,

$$p^2c^2 = (E-V_0)^2 - m^2c^4. \qquad (2.27)$$

The momentum is imaginary. There can only be exponentially decaying wave in the barrier. The calculated reflection coefficient is

$$R = 1. \qquad (2.28)$$

This is total reflection. Numerical results are depicted in figure 3.

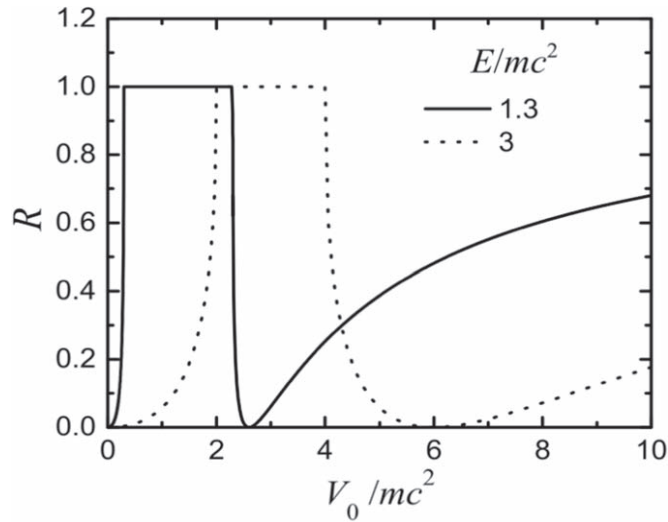

**Figure 3.** Reflection coefficient $R = |B|^2$ of one-dimensional Klein–Gordon equation in the potential barrier shown in figure 1. In the range $0 < V_0/mc^2 < 2E/mc^2$, the curve is symmetric with respect to $E/mc^2 = 1$. At $V_0 = 2E$ there is a transmission alley. When $V_0 \to \infty$, $R \to 1$.

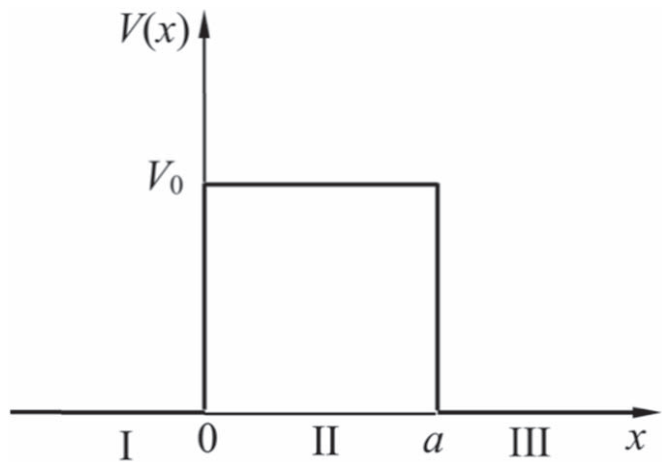

**Figure 4.** One-dimensional finitely wide potential barrier with width $a$ and height $V_0$.

The curves in figure 3 can be compared with those in figure 2. The qualitative behaviors are almost the same. Within the energy gap, the particle reflects totally. There exists a transmission alley around $V_0 = 2E$. The discrepancy is that in figure 3, as $V_0 \to \infty$, $R \to 1$, as can be seen from equation (2.26).

In summary, we have carefully treated the problem of potential barrier with infinite width. Both Dirac equation and Klein–Gordon equation are considered. In the latter case, decoupled Klein–Gordon equations are utilized.

## 3. Finitely wide potential barrier

The potential is of the form of figure 4.

$$V(x) = \begin{cases} 0, & x \leqslant 0 \\ V_0, & 0 < x \leqslant a, \\ 0, & x > a \end{cases} \quad (3.1)$$

A particle with energy $E$ and momentum $q$ is incident from $-\infty$ and moves rightwards. Let us investigate the solutions in both cases of Dirac equation and Klein–Gordon equation.

### 3.1. Dirac equation

In regions I and III where the potential is zero, the wave functions are respectively

$$\psi_I = \begin{pmatrix} qc \\ E - mc^2 \end{pmatrix} e^{iqx/\hbar} + B \begin{pmatrix} -qc \\ E - mc^2 \end{pmatrix} e^{-iqx/\hbar}, \; x < 0 \tag{3.2}$$

and

$$\psi_{III} = G \begin{pmatrix} qc \\ E - mc^2 \end{pmatrix} e^{iqx/\hbar}, \; x > a. \tag{3.3}$$

The energy-momentum relation in these two regions is

$$E^2 = q^2 c^2 + m^2 c^4. \tag{3.4}$$

In order to put down the wave function in region II, the barrier height $V_0$ has to be divided into four ranges as follows.

1. $V_0 < E - mc^2$

    In region II, $0 \leqslant x \leqslant a$ the wave function is

$$\psi_{II} = F_1 \begin{pmatrix} qc \\ E - V_0 - mc^2 \end{pmatrix} e^{ipx/\hbar} + F_2 \begin{pmatrix} -qc \\ E - V_0 - mc^2 \end{pmatrix} e^{-ipx/\hbar}, \; 0 \leqslant x \leqslant a, \tag{3.5}$$

The energy-momentum relation is

$$(E - V_0)^2 = p^2 c^2 + m^2 c^4. \tag{3.6}$$

We have simultaneous equations (3.2), (3.3) and (3.5). The boundary conditions are that at $x = 0$ and $x = a$ they should be continuous.

$$1 + B = F_1 + F_2. \tag{3.7}$$

$$F_1 e^{ipa/\hbar} + F_2 e^{-ipa/\hbar} = G e^{iqa/\hbar}. \tag{3.8}$$

$$q(1 - B) = p(F_1 + F_2). \tag{3.9}$$

$$p(F_1 e^{ipa/\hbar} - F_2 e^{-ipa/\hbar}) = qG e^{iqa/\hbar}. \tag{3.10}$$

The reflection amplitude is solved to be

$$B = -\frac{V_0 mc^2}{\zeta^2 - EV_0 + ipc\zeta \cot(pa/\hbar)}, \tag{3.11}$$

where $\zeta = \sqrt{E^2 - m^2 c^4}$. The reflection coefficient is $R = |B|^2$ and the transmission coefficient is $T = |G|^2 = 1 - R$. Numerical results are plotted by short-dashed line in figure 5. As the barrier height $V_0$ rises from 0 to $E - mc^2$, $R$ rises from 0 to 1 monotonically.

It is seen from (3.11) that as the barrier height is fixed, $R$ oscillates with the barrier width $a$.

$$\text{As } pa/\hbar = n\pi, \; R = 0. \tag{3.12}$$

This is resonant transmission. In this case, by equations (3.7)–(3.10) the coefficients in the wave functions can be solved: $B = 0$, $G = 1$, $F_1 = \frac{1}{2}\left(\frac{E - mc^2}{E - V_0 - mc^2} + \frac{q}{p}\right)$, $F_2 = \frac{1}{2}\left(\frac{E - mc^2}{E - V_0 - mc^2} - \frac{q}{p}\right)$.

2. $V_0 > E + mc^2$

    In region II, the wave function is in the form of

$$\psi_{II} = F_1 \begin{pmatrix} E - V_0 + mc^2 \\ pc \end{pmatrix} e^{ipx/\hbar} + F_2 \begin{pmatrix} E - V_0 + mc^2 \\ -pc \end{pmatrix} e^{-ipx/\hbar}, \; 0 < x < a, \tag{3.13}$$

where

$$(V_0 - E)^2 - m^2 c^4 = p^2 c^2. \tag{3.14}$$

The simultaneous equations now are (3.2), (3.3) and (3.13). By the conditions that the wave functions should be continuous at the boundaries $x = 0$ and $x = a$, the reflection amplitude is solved.

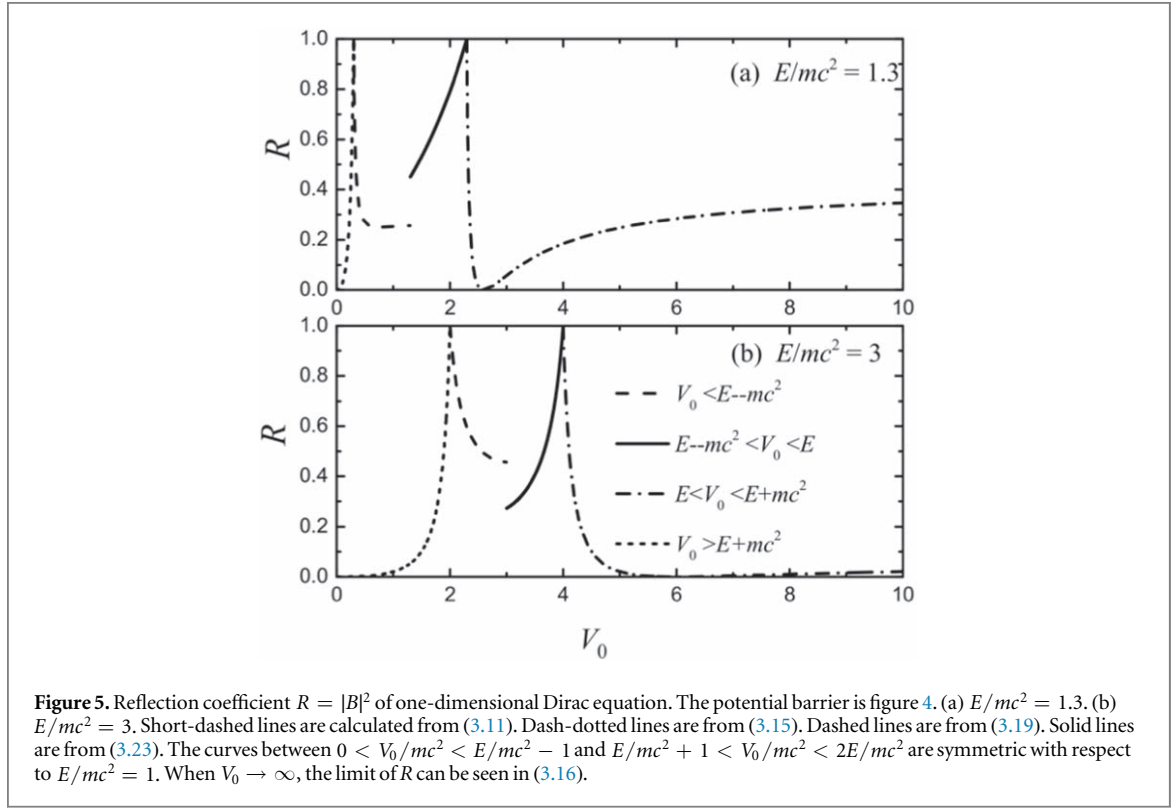

**Figure 5.** Reflection coefficient $R = |B|^2$ of one-dimensional Dirac equation. The potential barrier is figure 4. (a) $E/mc^2 = 1.3$. (b) $E/mc^2 = 3$. Short-dashed lines are calculated from (3.11). Dash-dotted lines are from (3.15). Dashed lines are from (3.19). Solid lines are from (3.23). The curves between $0 < V_0/mc^2 < E/mc^2 - 1$ and $E/mc^2 + 1 < V_0/mc^2 < 2E/mc^2$ are symmetric with respect to $E/mc^2 = 1$. When $V_0 \to \infty$, the limit of $R$ can be seen in (3.16).

$$B = -\frac{(V_0 - 2E)mc^2}{EV_0 - 2E^2 + \zeta^2 + ipc\zeta \cot(pa/\hbar)}. \tag{3.15}$$

The reflection coefficient is $R = |B|^2$ and transmission coefficient is $T = |G|^2 = 1 - R$.
In the ranges $0 < V_0 < E - mc^2$ and $E + mc^2 < V_0 < 2E$, there is a symmetry: the exchange $V_0 - E \Leftrightarrow E - V_0$ makes equation (3.11) become (3.15) and vice versa.
The $R$ versus $V_0$ is depicted by dash-dotted line in figure 5. As $V_0 = E + mc^2$ which means that the particle's momentum inside the barrier is zero, so that it has to reflect totally, $R = 1$. Then, with $V_0$ increases, $R$ decreases. At $V_0 = 2E$, it is again transmission alley and is a total transmission. After that, $R$ rises once more with increasing $V_0$. As $V_0 \to \infty$,

$$R \to \frac{m^2c^4}{E^2 + (E^2 - m^2c^4)\cot^2(pa/\hbar)} < 1. \tag{3.16}$$

This limit decreases with the energy $E$ increase. That is to say, the higher a particle's energy, the easier for it to transmit.
As the barrier height is fixed, the oscillation behavior of the reflection coefficient with the barrier width is still similar to that in equation (3.11). The conditions for resonant transmission are still (3.12). Under the resonant transmission, the coefficients in the wave functions can be solved: $B = 0$, $G = 1$, $F_1 = \frac{1}{2}\left(\frac{qc}{V_0 - E + mc^2} + \frac{pc}{E - mc^2}\right)$, $F_2 = \frac{1}{2}\left(\frac{qc}{V_0 - E + mc^2} - \frac{pc}{E - mc^2}\right)$.
Anyway, as $V_0 = 2E$, $p = q$. Therefore, the transmission alley always exists.
Now we are at the stage to consider the range

$$E - mc^2 < V_0 < E + mc^2. \tag{3.17}$$

We have analyzed in equations (2.14) and (2.15) that this means that the energy is in the relativistic gap and the momentum is an imaginary number. In the case of imaginary momentum, the wave functions were presented by equations (2.16) and (2.17).

   3. $E - mc^2 < V_0 < E$
     In region II the wave function should be in the form of equation (2.16),

$$\psi_{\mathrm{II}} = F_1 \begin{pmatrix} -ikc \\ E - V_0 - mc^2 \end{pmatrix} e^{kx/\hbar} + F_2 \begin{pmatrix} ikc \\ E - V_0 - mc^2 \end{pmatrix} e^{-kx/\hbar},\ 0 < x < a. \tag{3.18}$$

The simultaneous equations are (3.2), (3.3) and (3.18). By boundary conditions, the reflection amplitude is solved.

$$B = -\frac{V_0^2 - 2EV_0}{k^2c^2 - \zeta^2 - ikc\zeta\,\coth(ka/\hbar)}. \tag{3.19}$$

The reflection coefficient is $R = |B|^2$. Numerical results are the dashed line in figure 5. The two end points of the dashed line are at $V_0 = E - mc^2$ and $V_0 = E$. As $V_0 = E - mc^2$, $R = 1$, the total reflection, which is independent of the barrier width. As $V_0 = E$,

$$B = \frac{-E^2}{m^2c^4 - \zeta^2 - 2imc^2\zeta\,\cot(pa/\hbar)}. \tag{3.20}$$

What calculated by equation (3.20) is the lowest point of the dashed line.

Let us consider the variation of the reflection coefficient with the barrier width. Obviously, as $a = 0$, there is no potential barrier so that $B = 0$, i. e., the wave transmits totally. With the width $a$ increasing, the reflection coefficient rises. As $ka/\hbar \gg 1$, $\coth(ka/\hbar) = 1$, which results in

$$B = \frac{V_0^2 - 2EV_0}{k^2c^2 - \zeta^2 - ikc\zeta}. \tag{3.21}$$

In this case the reflection coefficient is calculated to be $R = 1$. This is just the result of infinitely wide barrier, see section 2.2.

4. $E < V_0 < E + mc^2$

In region II the wave function should be in the form of equation (2.17),

$$\psi_{\mathrm{II}} = F_1 \begin{pmatrix} E - V_0 + mc^2 \\ -ikc \end{pmatrix} e^{kx/\hbar} + F_2 \begin{pmatrix} E - V_0 + mc^2 \\ ikc \end{pmatrix} e^{-kx/\hbar},\ 0 < x < a. \tag{3.22}$$

The simultaneous equations are (3.2), (3.3) and (3.22). By the boundary conditions, we get

$$B = \frac{-V_0^2 mc^2}{\zeta^2 - V_0 E - i\sqrt{V_0^2 m^2 c^4 - (\zeta^2 - V_0 E)^2}\,\coth(pa/\hbar)}. \tag{3.23}$$

The reflection coefficient is $R = |B|^2$. Numerical results are depicted by the solid lines in figure 5. The two end points of the solid line are at $V_0 = E$ and $V_0 = E + mc^2$. As $V_0 = E + mc^2$, $R = 1$, the total reflection, which is independent of the barrier width. As $V_0 = E$,

$$B = \frac{E}{mc^2 - i\sqrt{E^2 - m^2c^4}\,\coth(ka/\hbar)}, \tag{3.24}$$

which gives the lowest point of solid lines in figure 5. As $V_0 = E + mc^2$, it is calculated that $R = 1$, and this total reflection is independent of barrier width.

Equations (3.20) and (3.23) have the same dependence on the barrier width. Hence, the dependences of the reflection coefficient on the barrier width are the same. As $a = 0$, it is a total transmission. With the width increasing, the reflection coefficient rises. As $ka/\hbar \gg 1$, $\coth(ka/\hbar) = 1$. This is a total reflection $R = 1$, which is the case of infinitely wide barrier, see section 2.2.

When the barrier height is within the range (3.17), the dashed and solid curves in figure 5 will become flat as the barrier width $a$ goes to infinity, just as the case in figure 2.

In figure 5, the reflection coefficient is discontinuous at $V_0 = E$, i. e., there is a jump. This jump will become zero as the barrier width goes to infinity.

In figure 5, in drawing dashed and solid lines, the barrier width is determined by $\coth(ka/\hbar) = 2$; in drawing short-dashed and dash-dotted lines, the width is determined by $\cot(ka/\hbar) = 1/2$. The two widths are different. Nevertheless, whatever the width is, the reflection coefficient is always 1 at the two positions $V_0 = E \pm mc^2$. Therefore, at these two points, the curves always connect. The dashed and solid curves are just to show the feature of reflection. In fact, for relativistic motion, the momentum is quite large and $ka/\hbar \gg 1$, so that $\coth(ka/\hbar) \approx 1$. Subsequently, the dashed and solid lines in figure 5 are actually very close to flat ones. That we choose $\coth(ka/\hbar) = 2$ is for the readers to see the details of the curves.

In any case, the reflection coefficient cannot be larger than 1.

Within relativistic energy gap, a particle's wave function can have exponential solutions. When Dirac proposed his equation, he solved the wave function of a free particle, but did not give the exponential solutions. This is because he solved in infinite space, so that the exponential solutions had to be zero. Here the potential barrier region is finite, so we can have nonzero exponential solutions. The exponential waves cause the reflection behavior as described by dashed and solid lines in figure 5.

### 3.2. Klein–Gordon equation

For the case of Klein–Gordon equation, the procedure is the same as in section 2.3.

In region I of $x < 0$ and region III of $x > a$, the wave functions are respectively

$$\psi_{\mathrm{I}} = e^{iqx/\hbar} + Be^{-iqx/\hbar} \qquad (3.25)$$

and

$$\psi_{\mathrm{III}} = Ge^{iqx/\hbar}. \qquad (3.26)$$

In these two regions, energy-momentum relation is

$$E^2 - m^2c^4 = q^2c^2. \qquad (3.27)$$

In region II of $0 \leqslant x \leqslant a$, the energy-momentum relation is

$$(E - V_0)^2 - m^2c^4 = q^2c^2. \qquad (3.28)$$

As $V_0 < E - mc^2$ and $V_0 > E + mc^2$, the wave function is

$$\psi_{\mathrm{III}} = F_1 e^{ipx/\hbar} + F_2 e^{-ipx/\hbar}. \qquad (3.29)$$

The boundary conditions are that at $x = 0$ and $x = a$, the wave functions and their derivatives are continuous. Thus, the reflection amplitude is solved to be

$$B = \frac{-p^2 + q^2}{p^2 + q^2 + 2ipq \cot(pa/\hbar)}. \qquad (3.30)$$

The reflection coefficient is $R = |B|^2$. As $V_0 \to \infty$, $R \to 1$, which is a feature similar to the case of low momentum motion.

From comparison of (3.30) and (3.11), it is seen that the dependence of the reflection coefficient on barrier width is similar. The resonant transmission condition is still (3.12). The reflection maxima are as follows.

$$\text{As } pa/\hbar = \left(n + \frac{1}{2}\right)\pi, R = \left(\frac{p^2 - q^2}{p^2 + q^2}\right)^2. \qquad (3.31)$$

As $E - mc^2 < V_0 < E + mc^2$, the momentum is imaginary. The wave function in region II is

$$\psi_{\mathrm{III}} = F_1 e^{kx/\hbar} + F_2 e^{-kx/\hbar}. \qquad (3.32)$$

The calculated reflection amplitude is

$$B = \frac{k^2 + q^2}{-k^2 + q^2 + 2ikq \coth(ka/\hbar)}. \qquad (3.33)$$

The reflection coefficient is $R = |B|^2$. In fact, if $k \to ip$ is taken, then equation (3.30) becomes (3.33). In this sense, these two equations are uniform. Of cause, their reflection behaviors differ greatly.

Numerical results are drawn in figure 6. All the features of the reflection coefficient are almost the same as those in figure 5. The only differences are that the curve is continuous at $V_0 = E$.

Both equations (3.30) and (3.33) remain unchanged under the exchange $V_0 - E \Leftrightarrow E - V_0$. Therefore, in figure 6, each curve within $0 < V_0/mc^2 < 2E/mc^2$ is symmetric with respect to $E/mc^2 = 1$.

### 3.3. Discussion of massless Dirac Fermions

It has been well known [15] that in metallic carbon nanotubes rolled up by graphene sheets, the quasi-particles near Fermi energy are zero-mass ones with spin-1/2. A remarkable feature for this kind of Dirac Fermion is their long mean free paths or high conductivity [16–21]. Theoretical calculation [22–24] showed the lack of back scattering in such systems. This kind of effect was explained by different reasons, such as long-ranged disorder potential [22] and the variation of Berry phase [23].

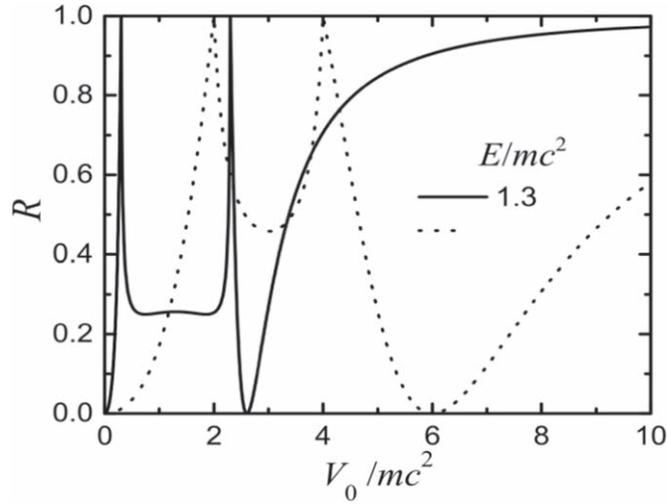

**Figure 6.** Reflection coefficient $R = |B|^2$ of one-dimensional decoupled Klein–Gordon equations. The potential barrier is figure 4. Solid and dotted lines are for $E/mc^2 = 1.3$ and 3, respectively. As $V_0 < E - mc^2$ and $V_0 > E + mc^2$, the curves are calculated by equation (3.30) where $\coth(ka/\hbar) = 2$ is taken; as $E - mc^2 < V_0 < E + mc^2$, the curves are calculated by equation (3.33) where $\cot(ka/\hbar) = 1/2$ is taken. When $V_0 \to \infty$, $R \to 1$.

According to the results in this paper, for a zero-mass Dirac Fermion, the transmissivity is always 1 when encountering a piecewise constant potential. This can be seen from equations (3.11) and (3.15) by letting the mass $m = 0$. Even for infinitely wide potential, equations (2.7), (2.8), (2.11) and (2.12) reveal that the refection coefficient will be zero if we take the mass $m = 0$. In the author's opinion, the reason of the total transmission is simply that the weights of the two components are the same. The discussion in the following two paragraphs defaults the mass $m = 0$. Consequently, the relativistic energy gap vanishes.

Let us first see the case of infinitely wide potential. The ratio of the amplitudes of the two components of the incident wave is 1:1, and that of the reflection wave is −1:1, see equation (2.4). In the potential region, the ratio is still 1:1, see equation (2.5) for $V_0 < E$ and (2.9) for $V_0 > E$. Since the wave function is continuous at the boundary, the appearance of the reflection wave will violate the continuity. Therefore, there is no way for the reflective wave to appear.

Then, we inspect the case of finitely wide barrier. It is seen from equations (3.2), (3.3), (3.5) and (3.13) that for the forward waves, the two amplitude ratios are always 1:1; while for the backward waves, the amplitude ratios are −1:1 for $V_0 < E$ and 1:−1 for $V_0 > E$. Let us look at the wave in region III where there is only transmission wave and its amplitude ratio is 1:1. The continuity of the wave function at the boundary $x = a$ prohibits the appearance of the backward wave in region II. Subsequently, at the boundary $x = 0$, the continuity condition in turn prohibits the appearance of the backward wave in region I. It is concluded that for massless Dirac Fermions, the reflection is always zero. This reasoning is valid for piecewise constant potentials.

As for a scalar potential varying smoothly, let us inspect the equations of the two components of Dirac equation.

$$-i\hbar c \varphi' = (E - V + mc^2)\chi. \tag{3.34a}$$

$$-i\hbar c \chi' = (E - V - mc^2)\varphi. \tag{3.34b}$$

As mass is zero, $m = 0$, an apparent solution is that

$$\chi = \pm \varphi. \tag{3.35}$$

It is easy to obtain the solution

$$\varphi(x) = \varphi(a) \exp\left[\pm \frac{i}{\hbar c}\left(E(x-a) - \int_a^x dx' V(x')\right)\right]. \tag{3.36}$$

For instance, if $V(x) = bx$, the solution will be $\varphi(x) = \varphi(a)\exp\left[\pm\frac{i}{\hbar c}\left(Ex - \frac{1}{2}bx^2\right)\right]$. Equation (3.36) discloses that the amplitude remains unchanged everywhere and equation (3.35) reveals that the amplitudes of the two components are the same. It is concluded that the reasoning for the piecewise constant scalar potential above is still valid.

For three-dimensional Dirac equation, the two spinor components satisfy

$$c\boldsymbol{\sigma} \cdot \boldsymbol{p}\, \varphi' = (E - V + mc^2)\chi \qquad (3.37a)$$

and

$$c\boldsymbol{\sigma} \cdot \boldsymbol{p}\, \chi' = (E - V - mc^2)\varphi. \qquad (3.37b)$$

As mass is zero, equation (3.35) still stands. That is to say, the amplitudes of the two components, which themselves are spinors, are the same.

As for two-dimensional case, when mass is zero, the two equations are

$$\hbar c(-i\partial_x + \partial_y)\varphi = (E - V)\chi \qquad (3.38a)$$

and

$$\hbar c(-i\partial_x - \partial_y)\chi = (E - V)\varphi. \qquad (3.38b)$$

Equation (3.38) cannot lead to (3.35). The equations were treated in [21], where it could be deduced that the asymmetry of the two equations of (3.38) caused a phase difference of the two components.

The discussions above demonstrate that the absolute ratios of the two components were 1 for massless Dirac Fermions. It was indeed so in some theoretical evaluations [15, 22, 25]. This provides a strong reason of the lack of back scattering. The massless Dirac Fermions move everywhere, which means a supercurrent.

It is seen that under a scalar potential Dirac Fermions are lack of back scattering. Then what about a vector potential? It was mentioned [22] that the back scattering would happen after a magnetic field was applied. We think that this is because a vector potential can make the amplitudes of the two components different. This is easily verified. Consider a two-dimensional plane. A magnetic field perpendicular to the plane is exerted with the vector potential being $\mathbf{A} = (-By, 0)$. We do not present the calculation details here. It can be calculated that the obtained eigenvalues are $E_{(\pm)} = \pm\sqrt{2(n+1)q\hbar cB}$ where $q$ is electric charge, the two components of which are the eigen functions of a harmonic oscillator with the subscripts differing by 1. Therefore, the two components are not the same. That is to say, a magnetic field makes the two components of the massless Dirac Fermion different. Consequently, the back scattering will occur.

It was mentioned [15] that for massless Dirac Fermions, for a barrier with width $a$, under the condition that $k_x a = n\pi$, there occurred total transmission. As a matter of fact, according to the results in this paper, this resonant condition also applies to massive ones and even for all particles such as Dirac Fermions and Klein–Gordon ones. In all these cases, once the condition equation (3.12) is met, the reflection will be zero. This is easily checked from equations (3.11), (3.15) and (3.30), as long as the potential is piecewise constants.

There are also particles having nonzero but very small masses. Neutrinos are such kind of particles. Since their masses are very small [26–28], they have very small reflection coefficients when encountering a scalar potential. This can be seen from figure 5 and equation (3.15). As $V_0 > 2E$, the reflection coefficient $R$ increases with $V_0$ from zero, and the highest value of $R$ can be seen from equation (3.16). The upper limit is $R < (mc^2/E)^2$. Obviously, the reflection is quite small. The deduction applies to the case of infinitely wide potential as well. The upper limit is also $R < (mc^2/E)^2$, see equation (2.13). When $V_0 < 2E$, for a finitely wide potential, there are dense points satisfying condition of resonant transmission $pa = n\pi$. For a scaler potential other than the piecewise constant one, equation (3.34) should be accounted for. When $mc^2/E \ll 1$, we can reasonably assume that $\varphi = \varphi_0 + \varphi_1$ and $\chi = \chi_0 + \chi_1$, where $\varphi_0$ and $\chi_0$ meet equation (3.35) as mass is zero. Then it is estimated that $|\varphi_1/\varphi_0| \sim mc^2/E$ and $|\chi_1/\chi_0| \sim mc^2/E$. That is to say, the amplitudes of the two components are quite close to each other. Consequently, the reflectivity should be very small.

## 4. Conclusions

In this work, the reflection coefficients of one-dimensional step potential barriers with infinite and finite widths are evaluated for a particle's relativistic stationary motion. The key is to distinguish the solving region as $E > V$ and $E < V$, where $E$ is the particle's energy and $V$ is potential. In each region, corresponding wave function is selected.

Klein's paradox of Dirac equation for spin-1/2 is figured out. The same paradox from Klein–Gordon equation for spin-0 is also solved. This equation is decoupled to be two, respectively applied to $E > V$ and $E < V$ regions, in the cases of piecewise constant potential.

In both cases of Dirac equation and Klein–Gordon equation, the reflection coefficient never exceeds 1. There is no need to produce electrons and positrons in vacuum, so that the concept of Dirac's Fermions Sea can be totally abandoned at least in the view of reflection from a potential barrier.

Both spin-1/2 and spin-0 particles are treated in a uniform way, and their reflection coefficient curves have the following same features.

There is an energy gap with width $2mc^2$ in the energy-momentum relation $(E - V)^2 = m^2c^4 + c^2p^2$. When a particle's energy is within this gap, its momentum is imaginary and its wave functions are of exponential forms.

There are two kinds of cases where the particle transmits totally. One is that the particle's energy is in transmission alley, $E = 2V$. The other is that when the barrier width meets the condition $pa = n\pi$, where $p$ is the particle's momentum in the barrier region, resonant transmission.

It can be concluded that for zero-mass spin-1/2 particles, such as the elementary excitations in graphenes, always transmit totally. For particles with very small masses, such as neutrinos, the transmission coefficient is nearly 1.

This research is supported by the National Key Research and Development Program of China [Grant No. 2016YFB0700102].

## ORCID iDs

Huai-Yu Wang  https://orcid.org/0000-0001-9107-6120

## References

[1] Klein O 1929 Die reflexion yon elektronen an einem Potentialsprung nach der relativistischen, Dynamik yon Dirac. Von *Z. Phys.* **53** 157
[2] Yndurain F J 1996 *Relativistic Quantum Mechanics and Introduction to Field Theory* (Berlin, Heidelberg: Springer) 87
[3] Holstein B R 1998 Klein's paradox *Am. J. Phys.* **66** 507
[4] Calogeracos A and Dombey M 1999 History and physics of the Klein paradox *Contem. Phys.* **40** 313
[5] Gerritsma R, Kirchmair G, Zahringer F, Solano E, Blatt R and Roos C F 2010 Quantum simulation of the Dirac equation *Nature* **463** 68
[6] Gerritsma R, Lanyon B P, Kirchmair G, Zahringer F, Hempel C, Casanova J, Garcia-Ripoll J J, Solano E, Blatt R and Roos C F 2011 Quantum simulation of the Klein paradox with trapped ions *Phys. Rev. Lett.* **106** 060503
[7] Umul Y Z 2019 A survey on Klein paradox *Optik* **181** 258
[8] Fuda M G and Furlani E 1982 Zitterbewegung and the Klein paradox for spin-zero particles *Am. J. Phys.* **50** 545
[9] Mahan G D 2005 *Quantum Mechanics in a Nutshell* (Princeton: Princeton University Press) 356
[10] Wachter A 2011 *Relativistic Quantum Mechanics* (Dordrecht: Springer Science + Business Media B. V.) 978-90-481-3644-5 (https://doi.org/10.1007/978-90-481-3645-2)
[11] Dombey M and Calogeracos A 1999 Seventy years of the Klein paradox *Phys. Rep.* **315** 41
[12] Greiner W, Muller B and Rafelski J 1985 *Chapter 5 in Quantum Electrodynamics of Strong Field* (Belin: Springer) 112–21
[13] Greiner W 2000 *Chapter 13 in Relativistic Quantum Mechanics-wave Equations* 3rd edn (Berlin: Springer) (http://www.fulviofrisone.com/attachments/article/466/Greiner%20W.%20Relativistic%20quantum%20mechanics.%20Wave%20equations%20(Springer,%202000).pdf)
[14] Huai-Yu W 2020 New results by low momentum approximation from relativistic quantum mechanics equations and suggestion of experiments *J. Phys. Commun.* **4** 125004
[15] McEuen P L, Bockrath M, Cobden D H, Yoon Y G and Louie S G 1999 Disorder, pseudospins, and backscattering in Carbon nanotubes *Phys. Rev. Lett.* **83** 5098
[16] Tans S J, Decoret M H, Dai H, Thess A, Smalley R E, Geerligs L J and Dekker C 1997 Individual single-wall carbon nanotubes as quantum wires *Nature* **386** 474
[17] Bockrath M, Cobden D H, McEuen P L, Chopra N G, Zettl A, Thess A and Smalley R E 1997 Single-electron transport in ropes of carbon nanotubes *Science* **275** 1922
[18] Frank S, Poncharal P, Wang Z L and de Heer W A 1998 Carbon nanotube quantum resistors *Science* **280** 1744
[19] Kasumov A Y, Deblock R, Kociak M, Reulet B, Bouchiat H, Khodos I I, Gorbatov Y B, Volkov V T, Journet C and Burghard M 1999 Supercurrents through single-walled carbon nanotubes *Science* **284** 1508
[20] Soh H T, Quate C F, Morpurgo A F, Marcus C M, Kong J and Dai H 1999 Integrated nanotube circuits: controlled growth and ohmic contacting of single-walled carbon nanotubes *Appl. Phys. Lett.* **75** 627
[21] Heersche H B, Pablo J H, Oostinga J B, Vandersypen L M K and Morpurgo A F 2007 Bipolar supercurrent in graphene *Nature* **446** 56
[22] Ando T and Nakanishi T 1998 Impurity scattering in carbon nanotubes—absence of back scattering—absence of back scattering *J. Phys. Soc. Japan* **67** 1704
[23] Ando T, Nakanishi T and Saito R 1998 Berry's phase and absence of back scattering in carbon nanotubes *J. Phys. Soc. Japan* **67** 2857
[24] Beenakker C W J 2008 Colloquium: andreev reflection and Klein tunneling in graphene *Rev. Mod. Phys.* **80** 1337
[25] Neto A H C, Guinea F, Peres N M R, Novoselov K S and Geim A K 2009 The electronic properties of graphene *Rev. Mod. Phys.* **81** 109
[26] Beringer J *et al* (Particle data group) review of particle physics *Phys. Rev.* D **86** 010001
[27] Fukugita M and Yanagida T 2003 *Physics of Neutrinos and Applications to Astrophysics* (Berlin: Springer) 267
[28] Zuber K 2012 *Neutrino Physics* 2nd edn (Boca Raton: CRC Press, Tayler & Francis Group) 139, 318, 388